\documentclass[11pt,a4paper]{article}
\usepackage{lmodern}
\usepackage[T1]{fontenc}
\usepackage[latin9]{inputenc}
\usepackage{amsmath}
\usepackage{amssymb}
\usepackage{esint}

\makeatletter

\pdfoutput=1 

\usepackage{jheppub}

\usepackage{etoolbox}
    \makeatletter
    \patchcmd{\maketitle}{\@fpheader}{}{}{}
    \makeatother

\usepackage{color}

\usepackage{amsfonts}

\usepackage{bbm}

\title{\boldmath Asymptotic symmetries in Carrollian theories of gravity with a negative cosmological constant}

\author{Alfredo P\'{e}rez}

\affiliation{Centro de Estudios Cient\'{i}ficos (CECs), Avenida Arturo Prat 514, Valdivia,
Chile.}

\affiliation{Facultad de Ingenier\'{i}a, Arquitectura y Dise\~no, Universidad San Sebasti\'{a}n, sede Valdivia, General Lagos 1163, Valdivia 5110693, Chile.}

\emailAdd{alfredo.perez@uss.cl}

\preprint{CECS-PHY-21/04}%

\abstract{
Asymptotic symmetries of electric and magnetic Carrollian gravitational
theories with a negative cosmological constant $\Lambda$ are analyzed
in 3+1 space-time dimensions. In the magnetic theory, the asymptotic symmetry algebra is given by the conformal Carroll
algebra in three dimensions, which is infinite-dimensional and isomorphic
to the BMS$_{4}$ algebra. These results are in full agreement with
holographic expectations, providing a new framework for the study of
Carrollian holography. On the contrary, in the case of the electric theory, the presence
of a negative $\Lambda$  turns out to be incompatible with a consistent
set of asymptotic conditions, that can be traced back to the absence
of a sensible ground state configuration. This can be
improved if the Carrollian theory obtained from an electric contraction
of Euclidean General Relativity is considered. In this case, asymptotic
conditions can be constructed with an asymptotic symmetry algebra given
by $so\left(1,4\right)$. However, it is
shown that the space of spherically symmetric solutions of this theory is degenerate.

}

\makeatother

\begin{document}
\maketitle

\newpage

\section{Introduction\label{sec:1 Introduction}}

Asymptotic symmetries are of fundamental importance in the description
of General Relativity. They define the physical symmetries of the
theory. In the presence of a negative cosmological constant, the asymptotic
symmetry algebra can be interpreted as a conformal symmetry acting
in one lower dimension, which is one of the cornerstones of the AdS/CFT
correspondence~\cite{Maldacena:1998re,Gubser:1998bc,Witten:1998qj}.
For instance, in four spacetime dimensions the asymptotic symmetry
algebra with Henneaux-Teitelboim boundary conditions~\cite{Henneaux:1985tv} is finite dimensional and given by the $AdS_{4}\simeq so(2,3)$
algebra, that can be regarded as a conformal
algebra acting on a three-dimensional spacetime.\footnote{In the asymptotic conditions
of Henneaux and Teitelboim~\cite{Henneaux:1985tv}, the 
boundary metric is fixed (Dirichlet boundary conditions) and is conformal to the three-dimensional Minkowski 
metric. Some generalizations that contain non-trivial fluxes 
 through the AdS boundary were considered in refs.~\cite{Compere:2019bua,Compere:2020lrt}
 leading to infinite-dimensional field-dependent algebras.} 

Recently, a new gravitational theory with a Carrollian structure was
introduced in ref.~\cite{Henneaux:2021yzg} as a particular ultrarelativistic
($c\rightarrow0$) contraction of General Relativity, known as ``Magnetic
Carroll gravity.'' A different Carrollian contraction of Einstein
gravity leads to the so-called ``Electric Carroll gravity'' that
was introduced long ago in the context of Strong Gravity~\cite{Isham:1975ur}
or as a zero signature limit of General Relativity~\cite{Teitelboim:1978wv,Henneaux:1979vn,Teitelboim:1981ua}.
 These Carrollian gravitational theories have a simpler structure than Einstein gravity  (the algebra of constraints and the
 constraints themselves acquire a simpler form in the Carroll limit),  and therefore they provide 
natural models to explore certain aspects of quantum gravity from a new perspective. In this sense, the study
of the symmetries of these theories acquires a particular importance. For a vanishing cosmological constant, the asymptotic symmetries
were recently studied in 3+1 space-time
dimensions in ref.~\cite{Perez:2021abf} for both contractions, magnetic and electric. 

It is then natural
to wonder if in the presence of a negative cosmological constant there
exists a similar interpretation as in the case of Einstein gravity,
i.e., that the asymptotic symmetries in the Carrollian gravitational
theories correspond to conformal Carrollian algebras in one lower
dimension. This question is particularly relevant in the context of holography.

Carroll symmetry was introduced independently by Levy-Leblond in ref.~\cite{levy1965nouvelle} and Sen Gupta in ref.~\cite{sen1966analogue}. Its Lie algebra was obtained from a contraction of the
Poincar\'e algebra in the limit when the speed of light vanishes, and
it has been used in diverse physical contexts. For example, in gravitation, electric Carroll gravity
turns out to be well suited to describe the behavior of spacetime near space-like
singularities along the lines of the Belinsky-Khalatnikov-Lifshitz (BKL) approach~\cite{Belinsky:1970ew,Belinsky:1982pk, henneaux1982quantification, 
Damour:2002et}. In this regime, time derivatives dominate over spatial 
gradients, which is precisely one of the main properties of Carrollian theories, the so-called ``ultra-locality.''
 This theory was also used as the starting point of an alternative perturbation 
 scheme for quantum gravity in terms of the signature parameter that mimicks the 
 quantization of a relativistic free particle~\cite{Teitelboim:1981ua, Henneaux:1981su,Teitelboim:1983fi}. Other recent applications of the Carroll symmetry 
  have been found  in the study of
   the symmetries of plane gravitational waves~\cite{Duval:2017els}, physics near black hole  horizons~\cite{Penna:2018gfx,Donnay:2019jiz}, 
   cosmology and dark matter~\cite{deBoer:2021jej}, fractons~\cite{Pena-Benitez:2021ipo,Casalbuoni:2021fel,Bidussi:2021nmp}, the asymptotic structure of 
   General Relativity near null infinity~\cite{Duval:2014uva, Ciambelli:2018wre,Herfray:2021qmp},  and a Carrollian description of 
   celestial holography~\cite{Donnay:2022aba}. Further applications of the Carroll symmetry can be found in refs.~\cite{dauutcourt1967characteristic,Dautcourt:1997hb,Duval:2014uoa,Bergshoeff:2014jla,Cardona:2016ytk,Bergshoeff:2016soe,Grumiller:2017sjh,Barducci:2018thr,Morand:2018tke,Bagchi:2019xfx,Ravera:2019ize,Ciambelli:2019lap,Gomis:2019nih,Bagchi:2019clu,Banerjee:2020qjj,Ammon:2020fxs,Grumiller:2020elf,Gomis:2020wxp,Bagchi:2021qfe,Rodriguez:2021tcz,Concha:2021jnn,Campoleoni:2021blr,Marsot:2021tvq,Figueroa-OFarrill:2021sxz,Chen:2021xkw,Bagchi:2022owq}.

In the classification of the possible ``kinematical
groups''  performed by Bacry and Levi-Leblond in ref.~\cite{Bacry:1968zf}, it was shown that there exists
an extension of the Carroll algebra that admits a non-vanishing cosmological
constant $\Lambda$. This cosmological extension of the Carroll algebra can be
obtained from an ultrarelativistic limit of the de Sitter or anti-de
Sitter algebras depending on the sign of $\Lambda$. In the 
particular case when the cosmological constant is negative, the ``Carroll AdS$_{4}$
algebra'' turns out to be isomorphic to the Poincar\'e algebra\footnote{In ref.~\cite{Bacry:1968zf}, this algebra was called ``Para-Poincar\'e.''}.

On the other hand, a remarkable relation between the conformal Carroll
algebra in three dimensions and the BMS$_{4}$ algebra was found in
ref.~\cite{Duval:2014uva}, where it was shown that both algebras
are isomorphic. The BMS$_{4}$ algebra is infinite-dimensional, and
originally appeared in the study of gravitational radiation in asymptotically
flat spacetimes at null infinity~\cite{Bondi:1962,Sachs:1962}. It
contains the Poincar\'e algebra, or equivalently the Carroll-AdS$_{4}$
algebra in our physical context, as a subalgebra.

In this article, it is shown that for magnetic Carroll gravity with
a negative cosmological constant in 3+1 space-time dimensions,
the asymptotic symmetry algebra is infinite-dimensional (in contrast
to the case in Einstein gravity with Henneaux-Teitelboim boundary conditions) and is given by the conformal Carroll
algebra in three dimensions, in agreement with the expectations coming
from holography. The possibility of having an infinite dimensional
algebra comes from the fact that there is a relaxation in the fall-off
of the fields as compared with the asymptotic behavior inherited from
General Relativity. This relaxation is possible because the algebra
of the constraints possesses an abelian subalgebra in the Carroll
limit. For a particular restriction of the asymptotic conditions,
the asymptotic symmetry algebra can be consistently truncated to the
finite dimensional ``Carroll-AdS$_{4}$ algebra.''

On the contrary, the case of electric Carroll gravity is radically
different. The presence of a negative cosmological constant turns
out to be incompatible with a consistent set of asymptotic conditions.
Indeed, the naive ground state configuration analogous to the AdS
spacetime in General Relativity is not a solution of the constraints
of the electric theory~\cite{Perez:2021abf,Hansen:2021fxi}. As a
consequence, it does not seems to be possible to construct a well-defined
set of asymptotic conditions with non-trivial conserved charges in
this case. However, if one considers the theory obtained from the
electric Carrollian contraction of Euclidean General Relativity with
$\Lambda<0$, then it is possible to find an alternative background
configuration that allows to construct a consistent set of asymptotic
conditions as deviations from it. In this case, the asymptotic symmetry
algebra is finite dimensional and given by the $so\left(1,4\right)$
algebra.


The plan of the paper is the following. In section~\ref{sec:Asymptotic-symmetries-in_Magnetic},
the asymptotic symmetries in magnetic Carroll gravity with $\Lambda<0$
are studied. After a brief review of the formulation of the theory,
the ground state solution obtained from a Carroll contraction of Anti-de
Sitter spacetime is introduced. This solution plays a fundamental
role in the construction of the asymptotic conditions, because it is
used as a background configuration. The proposed fall-off for the
fields is then exhibited, and the charges and asymptotic symmetries
are obtained. They are spanned by the three-dimensional conformal
Carroll algebra, which according to ref.~\cite{Duval:2014uva} is isomorphic
to BMS$_{4}$. Possible consistent truncations of the asymptotic symmetries
are also discussed. Finally, a solution of the magnetic theory that
resembles the Schwarzschild-AdS configuration is introduced and their
charges are computed.

In section~\ref{sec:Asymptotic-symmetries-in_Electric}, the asymptotic
structure of electric Carroll gravity with a negative cosmological
constant is analyzed. In particular, in subsection~\ref{subsec:Electric-theory-obtained}
it is shown that the configuration obtained from a Carroll contraction
of AdS spacetime is not a solution of this theory. The subsequent
problems for constructing consistent asymptotic conditions are then
discussed. In subsection~\ref{subsec:Electric-theory-Euclidean}
this situation is improved by considering the Carrollian theory obtained
from the electric contraction of Euclidean General Relativity with
$\Lambda<0$. Asymptotic conditions are then constructed, where the
symmetry algebra is canonically realized and spanned by $so(1,4)$
generators. The space of spherically symmetric solution of this theory
is also explored, showing that is degenerate. 

Finally, section~\ref{sec:5 Final-remarks} is devoted to the conclusions
and some final remarks.

\section{Asymptotic symmetries in magnetic Carroll gravity with a negative
cosmological constant\label{sec:Asymptotic-symmetries-in_Magnetic}}

\subsection{Hamiltonian formulation of magnetic Carroll gravity with a non-vanishing
cosmological constant}

Magnetic Carroll gravity was recently introduced by Henneaux and Salgado-Rebolledo
in ref.~\cite{Henneaux:2021yzg}. It is obtained from a ``magnetic
Carrollian contraction'' of General Relativity, and it was originally
formulated in Hamiltonian form\footnote{Recently, a covariant formulation of magnetic Carroll gravity was
provided in ref.~\cite{Hansen:2021fxi}. The covariant action was
obtained from an appropriate truncation of the next-to-leading order
term in a small speed of light expansion of General Relativity.}. In the presence of a cosmological constant $\Lambda$, the Hamiltonian
action reads
\begin{equation}
I=\int dtd^{3}x\,\left(\pi^{ij}\dot{g}_{ij}-N\mathcal{H}^{M}-N^{i}\mathcal{H}_{i}^{M}\right),\label{eq:action_principle_magnetic}
\end{equation}
where the constraint $\mathcal{H}^{M}$ and $\mathcal{H}_{i}^{M}$
are given by
\begin{align}
\mathcal{H}^{M} & =-\sqrt{g}\left(R-2\Lambda\right),\qquad\qquad\mathcal{H}_{i}^{M}=-2\pi_{i\mid j}^{\;j}.\label{eq:const_mag}
\end{align}
Here, the canonical variables are characterized by an Euclidean three-dimensional
metric $g_{ij}$, and their corresponding conjugate canonical momenta
$\pi^{ij}$ ($i,j=1,2,3$). They obey the following equal time Poisson
brackets
\[
\left\{ g_{ij}\left(x\right),\pi^{kl}\left(x'\right)\right\} =\frac{1}{2}\left(\delta_{i}^{k}\delta_{j}^{l}+\delta_{i}^{l}\delta_{j}^{k}\right)\delta\left(x,x'\right).
\]
The Lagrange multipliers are given by the lapse and shift functions
$N$, $N^{i}$, respectively. The symbol | denotes covariant differentiation
with respect to the metric $g_{ij}$, where $R$ is its Ricci scalar.

The time evolution of the canonical variables is described by Hamilton's
equations, which are directly obtained from the action principle~\eqref{eq:action_principle_magnetic}.
They read
\begin{equation}
\dot{g}_{ij}=N_{i\mid j}+N_{j\mid i}\,,\label{eq:Hamg}
\end{equation}
\begin{equation}
\dot{\pi}^{ij}=-N\sqrt{g}\left(R^{ij}-\frac{1}{2}g^{ij}R+\Lambda g^{ij}\right)+\sqrt{g}\left(N^{\mid i\mid j}-g^{ij}N_{\mid k}^{\,\mid k}\right)+\left(N^{k}\pi^{ij}\right)_{\mid k}-N_{\mid k}^{i}\pi^{kj}-N_{\mid k}^{j}\pi^{ki}\,.\label{eq:Hampi}
\end{equation}

The Carrollian structure of this theory is characterized by the algebra
of the first class constraints
\begin{align}
\left\{ \mathcal{H}^{M}\left(x\right),\mathcal{H}^{M}\left(x'\right)\right\}  & =0\,,\label{eq:HpHp}\\
\left\{ \mathcal{H}^{M}\left(x\right),\mathcal{H}_{i}^{M}\left(x'\right)\right\}  & =\mathcal{H}^{M}\left(x\right)\partial_{i}\delta\left(x,x'\right)\,,\label{eq:HpHi}\\
\left\{ \mathcal{H}_{i}^{M}\left(x\right),\mathcal{H}_{j}^{M}\left(x'\right)\right\}  & =\mathcal{H}_{i}^{M}\left(x'\right)\partial_{j}\delta\left(x,x'\right)+\mathcal{H}_{j}^{M}\left(x\right)\partial_{i}\delta\left(x,x'\right)\,,\label{eq:HiHj}
\end{align}
where the generators $\mathcal{H}^{M}\left(x\right)$ define an abelian
subalgebra, in contrast to the case of General Relativity. This particular
property is due to the fact that the Hamiltonian constraint $\mathcal{H}^{M}$
in~\eqref{eq:const_mag} does not depend on the canonical momenta.

The constraints $\mathcal{H}_{i}^{M}$ generate coordinate transformations
on the hypersurfaces defined at slices of constant time, while $\mathcal{H}^{M}$
generates normal surface deformations. The smeared form of the generators
of arbitrary surface deformations then takes the form
\begin{equation}
G\left[\xi,\xi^{i}\right]=\int d^{3}x\left(\xi\,\mathcal{H}^{M}+\xi^{i}\,\mathcal{H}_{i}^{M}\right)+Q_{M}\,,\label{eq:Generators}
\end{equation}
where $\xi$ and $\xi^{i}$ correspond to the parameters associated
with normal and tangential deformations, respectively.

Following the
Regge-Teitelboim approach \cite{Regge:1974zd}, $Q_{M}$ is a surface
term that must be added to ensure that the generators~\eqref{eq:Generators}
possess well-defined functional derivatives. This means that all the boundary terms 
that are obtained from the variation of the bulk part of the generator must be cancelled with
the variation of the surface term $Q_{M}$. Therefore, if the generator of an improper (large) gauge transformation
 is evaluated on a given solution, the bulk part vanishes because is linear in the constraints, and 
only the boundary term remains, defining the corresponding conserved charge associated
with this symmetry~\cite{Benguria:1976in}.

From the form of the constraints in eq.~\eqref{eq:const_mag} one finds
\begin{equation}
\delta Q_{M}=\oint d^{2}s_{l}\left[G^{ijkl}\left(\xi\delta g_{ij\mid k}-\xi_{\mid k}\delta g_{ij}\right)+2\xi_{k}\delta\pi^{kl}+\left(2\xi^{k}\pi^{jl}-\xi^{l}\pi^{jk}\right)\delta g_{jk}\right]\,.\label{eq:deltaQ}
\end{equation}
Here $G^{ijkl}$ denotes the inverse of the de Witt supermetric, given
by
\[
G^{ijkl}=\frac{1}{2}\sqrt{g}\left(g^{ik}g^{jl}+g^{il}g^{jk}-2g^{ij}g^{kl}\right)\,.
\]

The transformation laws of the canonical variables are obtained by
acting with $G\left[\xi,\xi^{i}\right]$ on them 
\[
\delta g_{ij}=\left\{ g_{ij},G\left[\xi,\xi^{k}\right]\right\} ,\qquad\delta\pi^{ij}=\left\{ \pi^{ij},G\left[\xi,\xi^{k}\right]\right\} .
\]
Explicitly, they read
\begin{equation}
\delta g_{ij}=\xi_{i\mid j}+\xi_{j\mid i}\,,\label{eq:transfg}
\end{equation}
\begin{equation}
\delta\pi^{ij}=-\xi\sqrt{g}\left(R^{ij}-\frac{1}{2}g^{ij}R+\Lambda g^{ij}\right)+\sqrt{g}\left(\xi^{\mid i\mid j}-g^{ij}\xi_{\mid k}^{\,\mid k}\right)+\left(\xi^{k}\pi^{ij}\right)_{\mid k}-\xi_{\mid k}^{i}\pi^{kj}-\xi_{\mid k}^{j}\pi^{ki}\,.\label{eq:transfpi}
\end{equation}

\subsection{Magnetic Carrollian ground state}

In order to construct a consistent set of asymptotic conditions, generically
one must consider deviations of the fields with respect to a certain
background configuration in the asymptotic expansion. For instance,
in the case of General Relativity with a negative cosmological constant,
the corresponding background is Anti-de Sitter spacetime, with isometries
that are spanned by the AdS$_{4}$ algebra.

For magnetic Carroll gravity, the natural background is given by the
solution obtained from a Carroll contraction of Anti-de Sitter spacetime.
It is given by
\begin{equation}
\bar{g}_{ij}dx^{i}dx^{j}=\frac{dr^{2}}{\left(\frac{r^{2}}{l^{2}}+1\right)}+r^{2}\gamma_{AB}dx^{A}dx^{B}\,,\qquad\bar{\pi}^{ij}=0\,,\label{eq:background}
\end{equation}
\[
\bar{N}=\sqrt{\frac{r^{2}}{l^{2}}+1}\,,\qquad\bar{N}^{i}=0.
\]
Here $\Lambda=-3/l^{2}$, and 
\[
\gamma_{AB}dx^{A}dx^{B}=d\theta^{2}+\sin^{2}\theta d\phi^{2},
\]
is the metric of the round 2-sphere. Its determinant
is denoted by $\gamma$. In what follows the indices $A,B=1,2$ are lowered
and raised with this metric. 

Note that the constraints~\eqref{eq:const_mag} are automatically
fulfilled because the spatial metric is of constant curvature, and
$\bar{\pi}^{ij}$ vanishes. In ref.~\cite{Figueroa-OFarrill:2019sex},
this solution was obtained from the quotient of the Poincar\'e group
by the subgroup formed by spatial rotations and translations. It defines
an homogeneous space with symmetries given by the Carroll-AdS$_{4}$
algebra. 

To determine the Carrollian isometries of the solution in the context
of this particular theory, one can use the transformation laws of
the canonical variables in eqs.~\eqref{eq:transfg} and~\eqref{eq:transfpi}.
Requiring that $\delta\bar{g}_{ij}=\delta\bar{\pi}^{ij}=0$, one finds
that the following equations must be obeyed by the parameters $\xi$
and $\xi^{i}$
\begin{equation}
\xi_{i\mid j}+\xi_{j\mid i}=0\,,\qquad\xi_{\mid i\mid j}-\bar{g}_{ij}\xi_{\mid k}^{\,\mid k}+\frac{2\xi}{l^{2}}\bar{g}_{ij}=0\,.\label{eq:Hamg-1}
\end{equation}
The explicit solution is given by
\[
\xi=r\left(\vec{\beta}\cdot\hat{r}\right)+T\sqrt{\frac{r^{2}}{l^{2}}+1}\,,\qquad\xi^{r}=\left(\vec{\alpha}\cdot\hat{r}\right)\sqrt{\frac{r^{2}}{l^{2}}+1}\,,
\]
\[
\xi^{A}=\frac{\epsilon^{AB}}{\sqrt{\gamma}}\partial_{B}\left(\vec{\omega}\cdot\hat{r}\right)+\frac{\partial^{A}\left(\vec{\alpha}\cdot\hat{r}\right)}{r}\sqrt{\frac{r^{2}}{l^{2}}+1}\,.
\]
Here, the constant $T$ is the parameter of time translations, and
$\vec{\alpha}$, $\vec{\beta}$, $\vec{\omega}$ are three-vectors
associated with spatial translations, Carrollian boosts and spatial
rotations, respectively. The vector $\hat{r}=\left(\sin\theta\cos\phi,\sin\theta\sin\phi,\cos\theta\right)$
denotes the unit normal to the two-sphere.

The algebra of Carrollian isometries can be obtained from the composition
law of the parameters derived from the ``Carrollian surface deformation
algebra'' in eqs.~\eqref{eq:HpHp}-\eqref{eq:HiHj} 

\begin{equation}
\xi_{3}^{\perp}=\xi_{1}^{i}\partial_{i}\xi_{2}^{\perp}-\xi_{2}^{i}\partial_{i}\xi_{1}^{\perp}\,,\qquad\xi_{3}^{i}=\xi_{1}^{j}\partial_{j}\xi_{2}^{i}-\xi_{2}^{j}\partial_{j}\xi_{1}^{i}\,.\label{eq:comp_Carroll}
\end{equation}
It is given by 
\[
\beta_{3}^{K}=\frac{1}{l^{2}}\left(T_{2}\alpha_{1}^{K}-T_{1}\alpha_{2}^{K}\right)+\epsilon_{IJK}\left(\beta_{2}^{I}\omega_{1}^{J}-\beta_{1}^{I}\omega_{2}^{J}\right)\,,\qquad T_{3}=\delta_{IJ}\left(\alpha_{1}^{I}\beta_{2}^{J}-\alpha_{2}^{I}\beta_{1}^{J}\right),
\]
\[
\alpha_{3}^{K}=-\epsilon_{IJK}\left(\alpha_{1}^{I}\omega_{2}^{J}-\alpha_{2}^{I}\omega_{1}^{J}\right)\,,\qquad\omega_{3}^{K}=\frac{1}{l^{2}}\epsilon_{IJK}\alpha_{1}^{I}\alpha_{2}^{J}-\epsilon_{IJK}\omega_{1}^{I}\omega_{2}^{J},
\]
where $I,J,K=1,2,3$ label the indices associated with the three-vectors. 

This composition law defines the Carroll-AdS$_{4}$ algebra, which
possesses the following non-vanishing commutators
\begin{equation}
\left\{ J_{I},J_{J}\right\} =-\epsilon_{IJK}J_{K}\,,\qquad\left\{ P_{I},J_{J}\right\} =-\epsilon_{IJK}P_{K}\,,\qquad\left\{ K_{I},J_{J}\right\} =-\epsilon_{IJK}K_{K}\,,\label{eq:Carroll_AdS_1}
\end{equation}
\begin{equation}
\left\{ P_{I},K_{J}\right\} =\delta_{IJ}E\,,\qquad\left\{ P_{I},P_{J}\right\} =\frac{1}{l^{2}}\epsilon_{IJK}J_{K}\,,\qquad\left\{ P_{I},E\right\} =\frac{1}{l^{2}}K_{I}\,.\label{eq:Carroll_AdS_2}
\end{equation}
Here $\vec{J}$, $\vec{K}$, $\vec{P}$ are the generators associated
with spatial rotations, Carrollian boosts and spatial translations,
respectively. The generator associated with time translations, i.e.,
the Carrollian energy, is denoted by $E$.

Note that the equivalence with the Poincar\'e algebra becomes explicit
if we change $P_{I}\rightarrow\frac{1}{l}K_{I}$ and $K_{I}\rightarrow lP_{I}$.
When $l\rightarrow\infty$ one recovers the original Carroll algebra
of Levy-Leblond~\cite{levy1965nouvelle}.

\subsection{Asymptotic conditions and symmetry algebra}

\subsubsection{Fall-off of the fields}

A key step in the process of finding the asymptotic symmetries is
to specify the behavior of the fields, in this case the canonical
variables $g_{ij}$, $\pi^{ij}$, in the large $r$ expansion. The
fall-off of the fields must guarantee that the symplectic term in
the action~\eqref{eq:action_principle_magnetic} is finite and that
the charges are finite and integrable in the functional sense\footnote{We are assuming 
that there are no fluxes across the boundary.}. For
this purpose, we will consider deviations with respect the background
configuration defined in eq.~\eqref{eq:background}. The proposed
asymptotic conditions are given by

\begin{align}
g_{rr} & =\frac{l^{2}}{r^{2}}-\frac{l^{4}}{r^{4}}+\frac{f_{rr}}{r^{5}}+\mathcal{O}\left(r^{-6}\right),\label{eq:grr}\\
g_{rA} & =\frac{f_{rA}}{r^{4}}+\mathcal{O}\left(r^{-5}\right),\\
g_{AB} & =r^{2}\gamma_{AB}+h_{AB}+\frac{f_{AB}}{r}+\mathcal{O}\left(r^{-2}\right),
\end{align}
\begin{align}
\pi^{rr} & =\frac{p^{rr}}{r}+\mathcal{O}\left(r^{-2}\right),\\
\pi^{rA} & =-\frac{D_{B}\tilde{k}_{\left(2\right)}^{AB}}{r}+\frac{p^{rA}}{r^{2}}+\mathcal{O}\left(r^{-3}\right),\\
\pi^{AB} & =\frac{\tilde{k}_{\left(2\right)}^{AB}}{r^{2}}+\frac{k_{\left(4\right)}^{AB}}{r^{4}}+\frac{p^{AB}}{r^{5}}+\mathcal{O}\left(r^{-6}\right).\label{eq:piab}
\end{align}
Here, $D_{A}$ is the covariant derivative associated with the metric
$\gamma_{AB}$. For a tensor $X^{AB}$, its trace is denoted by $\tilde{X}:=X^{AB}\gamma_{AB}$
and its traceless part by $\tilde{X}^{AB}:=X^{AB}-\frac{1}{2}\gamma^{AB}\tilde{X}$
.

It is worth pointing out that the terms with $h_{AB}$, $\tilde{k}_{\left(2\right)}^{AB}$
and $k_{\left(4\right)}^{AB}$ are not present in the fall-off of
the fields inherited from General Relativity. Indeed, this ``relaxation''
in the asymptotic behavior of the fields is of fundamental importance
for the existence of an infinite-dimensional asymptotic symmetry algebra,
and has its origin in the abelian subalgebra~\eqref{eq:HpHp} of the
Carrollian surface deformation algebra.

The asymptotic conditions~\eqref{eq:grr}-\eqref{eq:piab} are preserved
by parameters of the form

\begin{align}
\xi & =\frac{r}{l}\,T\left(\theta,\phi\right)+\frac{l\left(\Delta+2\right)T\left(\theta,\phi\right)}{4r}+\dots\,,\label{eq:param1}\\
\xi^{r} & =-\frac{r}{2}D_{A}Y^{A}\left(\theta,\phi\right)-\frac{l^{2}}{4}D_{A}Y^{A}\left(\theta,\phi\right)\frac{1}{r}+\dots\,,\\
\xi^{A} & =Y^{A}\left(\theta,\phi\right)-\frac{l^{2}}{4r^{2}}D^{A}D_{B}Y^{B}\left(\theta,\phi\right)+\dots\,.\label{eq:param3}
\end{align}
Here, $\Delta=D^{A}D_{A}$, and $T\left(\theta,\phi\right)$ is an
arbitrary scalar function on the sphere. The vector $Y^{A}$ is constrained
to obey the two-dimensional conformal Killing equation
\[
D_{A}Y_{B}+D_{B}Y_{A}-\gamma_{AB}D_{C}Y^{C}=0\,.
\]

\subsubsection{Conserved charges and asymptotic symmetry algebra}

The canonical generators of the asymptotic symmetries parametrized
by $T$ and $Y^{A}$ can be directly obtained by evaluating the surface
term $\delta Q_{M}$ in eq.~\eqref{eq:deltaQ}. If the parameters
are assumed to have a vanishing functional variation, the charge can
be readily integrated and yields
\begin{align}
Q_{M} & =\oint d^{2}x\sqrt{\gamma}\left(T\,\mathcal{P}+Y^{A}\,\mathcal{J}_{A}\right),\label{eq:charge_mag}
\end{align}
where
\begin{equation}
\mathcal{P}:=\frac{1}{l^{2}}\left(3\tilde{f}+\frac{2}{l^{2}}f_{rr}\right)\,,\qquad\mathcal{J}_{A}:=2\gamma^{-\frac{1}{2}}\gamma_{AB}p^{rB}.\label{eq:calP}
\end{equation}

If $\lambda=(T,Y^{A})$ denotes the parameters, the algebra can be
obtained from the identity $\delta_{2}Q\left[\lambda_{1}\right]=\left\{ Q\left[\lambda_{1}\right],Q\left[\lambda_{2}\right]\right\} =Q\left[\lambda_{3}\right]$.
Using the transformation laws for the fields $\mathcal{P}$ and $\mathcal{J}_{A}$
given by
\begin{align*}
\delta\mathcal{P} & =Y^{A}\partial_{A}\mathcal{P}+\frac{3}{2}\mathcal{P}\,D_{A}Y^{A}\,,\\
\delta\mathcal{J}_{A} & =\frac{1}{2}T\partial_{A}\mathcal{P}+\frac{3}{2}\mathcal{P}\partial_{A}T+Y^{B}D_{B}\mathcal{J}_{A}+\mathcal{J}_{B}D_{A}Y^{B}+\mathcal{J}_{A}D_{B}Y^{B}+\frac{3}{l^{2}}D^{B}\left(T\tilde{f}_{AB}\right)\,,
\end{align*}
it can be shown that the algebra closes according to
\begin{align}
T_{3} & =Y_{1}^{A}\partial_{A}T_{2}-Y_{2}^{A}\partial_{A}T_{1}+\frac{1}{2}\left(T_{1}D_{A}Y_{2}^{A}-T_{2}D_{A}Y_{1}^{A}\right),\label{eq:comp1}\\
Y_{3}^{A} & =Y_{1}^{C}\partial_{C}Y_{2}^{A}-Y_{2}^{C}\partial_{C}Y_{1}^{A}.\label{eq:comp2}
\end{align}
This is precisely the composition law of the BMS$_{4}$ algebra (see
e.g.~\cite{Barnich:2009se}). Therefore, using the results of~\cite{Duval:2014uva},
it can be concluded that the asymptotic symmetry algebra of magnetic
Carroll gravity with $\Lambda<0$ is given by the infinite-dimensional
conformal Carroll algebra in three-dimensions, in full agreement with
the expectations coming from holography.

Note that the canonical realization of the algebra does not contain
central charges, and also admits ``superrotations'' in the sense
of ref.~\cite{Barnich:2009se} if the conformal Killing vectors $Y^{A}$
are locally-defined. Alternatively, the composition laws~\eqref{eq:comp1}
and~\eqref{eq:comp2} can also be obtained using the asymptotic form
of the parameters~\eqref{eq:param1}-\eqref{eq:param3} in eq.~\eqref{eq:comp_Carroll}.

If the parameters $T$, $Y^{A}$ are defined globally, they can be
expanded in spherical harmonics according to
\[
Y^{A}=\partial^{A}\left(\frac{\vec{\alpha}}{l}\cdot\hat{r}\right)+\frac{\epsilon^{AB}}{\sqrt{\gamma}}\partial_{B}\left(\vec{\omega}\cdot\hat{r}\right)\,,\qquad T=T_{0}+l\vec{\beta}\cdot\hat{r}+\sum_{\text{\ensuremath{\ell=2}}}^{\infty}\sum_{m=-\ell}^{\ell}T_{\ell,m}Y_{\ell,m}\,.
\]
Analogously, the fields $\mathcal{P}$ and $\mathcal{J}_{A}$ can
be expanded as follows
\begin{equation}
\mathcal{J}_{A}=\frac{3l}{8\pi}\partial_{A}\left(\vec{P}\cdot\hat{r}\right)+\frac{3}{8\pi}\sqrt{\gamma}\epsilon_{AB}\partial^{B}\left(\vec{J}\cdot\hat{r}\right)\,,\qquad\mathcal{P}=\frac{E}{4\pi}+\frac{3}{4\pi l}\vec{K}\cdot\hat{r}+\sum_{\text{\ensuremath{\ell=2}}}^{\infty}\sum_{m=-\ell}^{\ell}P_{\ell,m}Y_{\ell,m}\,.\label{eq:JP}
\end{equation}
Therefore, the charge~\eqref{eq:charge_mag} can be written as 
\begin{align*}
Q_{M} & =T_{0}E+\vec{\alpha}\cdot\vec{P}+\vec{\beta}\cdot\vec{K}+\vec{\omega}\cdot\vec{J}+\sum_{\text{\ensuremath{\ell=2}}}^{\infty}\sum_{m=-\ell}^{\ell}T_{\ell,m}P_{\ell,m}\,.
\end{align*}
The subset of generators $E$, $\vec{P}$, $\vec{K}$ and $\vec{J}$
defines a Carroll-AdS$_{4}$ subalgebra, with Poisson brackets that
close according to eqs.~\eqref{eq:Carroll_AdS_1} and~\eqref{eq:Carroll_AdS_2}.
The additional infinite number of generators $P_{\ell,m}$ define the
supertranslations.

In this analysis, the canonical generators of the BMS$_{4}$ algebra
are well-defined, in contrast with the case of asymptotically flat
spacetimes at null infinity. In that context, there appear non-integrable
terms in the charges associated with the flux of gravitational radiation
across future/past null infinity~\cite{Barnich:2011mi} (see also~\cite{Bunster:2018yjr}).

\subsubsection{Consistent truncations of the asymptotic conditions}

Some consistent truncations of the asymptotic conditions \eqref{eq:grr}-\eqref{eq:piab}
are possible. From the transformation law of $\tilde{h}_{AB}$ written
in stereographic coordinates $z=e^{i\phi}\cot\left(\theta/2\right)$
\[
\delta h_{zz}=Y^{\bar{z}}\partial_{\bar{z}}h_{zz}+Y^{z}\partial_{z}h_{zz}+2\left(\partial_{z}Y^{z}\right)h_{zz}-\frac{1}{2}l^{2}\partial_{z}^{3}Y^{z},
\]
it is clear that if the condition $\tilde{h}_{AB}=0$ is imposed,
then $\partial_{z}^{3}Y^{z}=\partial_{\bar{z}}^{3}Y^{\bar{z}}=0$,
restrictions that eliminate the possibility of having superrotations.

From the transformation law of $\tilde{k}_{\left(2\right)}^{AB}$
\[
\delta\tilde{k}_{\left(2\right)}^{AB}=-\frac{\sqrt{\gamma}T}{l^{2}}\tilde{h}^{AB}+\sqrt{\gamma}\left(D^{A}D^{B}-\frac{1}{2}\gamma^{AB}\Delta\right)T+\frac{1}{2}\left(D_{C}Y^{C}\right)\tilde{k}_{\left(2\right)}^{AB}+\mathcal{L}_{Y}\tilde{k}_{\left(2\right)}^{AB},
\]
if we set
\begin{equation}
\tilde{h}_{AB}=0\,,\qquad\tilde{k}_{\left(2\right)}^{AB}=0\,,\label{eq:trunc}
\end{equation}
 the following condition on the parameter $T\left(\theta,\phi\right)$
is obtained
\[
\left(D^{A}D^{B}-\frac{1}{2}\gamma^{AB}\Delta\right)T=0\,.
\]
This condition annihilates the modes with $\ell\geq2$ in the spherical
harmonics expansion. Thus,
\[
T=T_{0}+l\vec{\beta}\cdot\hat{r}\,.
\]
Consequently, the modes of $\mathcal{P}\left(\theta,\phi\right)$
that contribute to the charges are only those with $\ell=0$ and $\ell=1$,
i.e., $E$ and $\vec{K}$. Therefore, the conditions~\eqref{eq:trunc}
truncate the infinite-dimensional BMS$_{4}$ algebra to the finite-dimensional
Carroll-AdS$_{4}$ algebra. With these restrictions (together with
$\tilde{k}_{\left(4\right)}^{AB}=0$, that does not play any fundamental
role), the fall-off in eqs.~\eqref{eq:grr}-\eqref{eq:piab} reduces
to a truncated set of asymptotic conditions with the same fall-off
as the one used in General Relativity with a negative cosmological
constant~\cite{Henneaux:1985tv}.

On the other hand, it is a well known fact that there is no a unique
embedding of the Poincar\'e/Carroll-AdS$_{4}$ algebra within the BMS$_{4}$
algebra~\cite{Sachs:1962}. However, the truncation of the asymptotic
conditions characterized by~\eqref{eq:trunc}, selects a ``preferred''
Poincar\'e/Carroll-AdS$_{4}$ algebra that corresponds to the algebra
that is inherited from General Relativity in the Carroll limit.

\subsection{Carrollian Schwarzschild-AdS solution}

Let us consider the solution of magnetic Carroll gravity with a negative
cosmological constant whose spatial metric coincides with that of
the Schwarzschild-AdS solution in General Relativity

\begin{equation}
g_{ij}dx^{i}dx^{j}=\frac{dr^{2}}{\left(\frac{r^{2}}{l^{2}}+1-\frac{M}{8\pi r}\right)}+r^{2}\left(d\theta^{2}+\sin^{2}\theta d\phi^{2}\right)\,,\qquad\pi^{ij}=0\,.\label{eq:mag_Schw_AdS}
\end{equation}
The lapse and shift functions are given by 
\begin{equation}
N=\sqrt{\frac{r^{2}}{l^{2}}+1-\frac{M}{8\pi r}}\,,\qquad N^{i}=0\,.\label{eq:lsmag_Schw_AdS}
\end{equation}
This solution was discussed using isotropic coordinates in a covariant
formulation of the magnetic theory in ref.~\cite{Hansen:2021fxi}.

From the asymptotic expansion of the spatial metric in eq.~\eqref{eq:mag_Schw_AdS}
\begin{align*}
g_{rr} & =\frac{l^{2}}{r^{2}}-\frac{l^{4}}{r^{4}}+\frac{Ml^{4}}{8\pi r^{5}}+\mathcal{O}\left(r^{-6}\right)\,,\\
g_{rA} & =0\,,\\
g_{AB} & =r^{2}\gamma_{AB}\,,
\end{align*}
 it is clear that fits within the asymptotic conditions~\eqref{eq:grr}-\eqref{eq:piab}
with

\[
f_{rr}=\frac{l^{4}}{8\pi}M.
\]
Therefore, from~\eqref{eq:calP} one finds

\[
\mathcal{P}=\frac{1}{4\pi}M.
\]
For this solution, $\mathcal{P}$ contains only the zero mode in the
expansion in spherical harmonics. Consequently, according to~\eqref{eq:calP},
the only non-trivial charge is the Carroll energy, given by

\[
E=M\,.
\]

Note that, as in the case with $\Lambda=0$ discussed in~\cite{Perez:2021abf},
the canonical variables, as well as the lapse and the shift in eqs.~\eqref{eq:mag_Schw_AdS},~\eqref{eq:lsmag_Schw_AdS} exactly coincide
with those of the Schwarzschild-AdS solution in General Relativity.
In spite of the fact that in the Carrollian theory there is no a notion
of a four-dimensional Riemannian metric that can be reconstructed
from them, in ref.~\cite{Perez-Troncoso-2022} it is shown that the
solution in~\eqref{eq:mag_Schw_AdS} and~\eqref{eq:lsmag_Schw_AdS}
can be interpreted as a ``Carrollian black hole'' when appropriate
regularity conditions are imposed on the ``thermal'' Carrollian
geometry.

\section{Asymptotic symmetries in electric Carroll gravity with a negative
cosmological constant\label{sec:Asymptotic-symmetries-in_Electric}}

The gravitational theory obtained from the electric Carroll contraction
of General Relativity was originally introduced in the 70's as a strong
coupling limit of Einstein gravity~\cite{Isham:1975ur}, or alternatively
as a ``zero signature limit'' of it~\cite{Teitelboim:1978wv,Henneaux:1979vn,Teitelboim:1981ua}.
In this limit, the Hamiltonian constraint does not depend on derivatives
of the spatial metric, acquiring only a quadratic dependence on the
momenta. This particular form of the Hamiltonian resembles that of
a massive relativistic free particle, where the role of the mass is
played by the cosmological constant. This similarity was used in ref.~\cite{Teitelboim:1981ua} as a starting point of an alternative perturbative
approach to quantum gravity. In ref.~\cite{Henneaux:1979vn}, an action
for electric Carroll gravity that is manifestly invariant under Carrollian
changes of coordinates, was introduced. By virtue of their ultra-local
properties (neighboring points are causally disconnected), this theory
turns out to be useful to describe the properties of spacetime near
space-like singularities, along the lines of the Belinsky-Khalatnikov-Lifshitz
approach~\cite{Belinsky:1970ew,Belinsky:1982pk,henneaux1982quantification,Damour:2002et}.

Recently, an analysis of the asymptotic structure of electric Carroll
gravity was performed in the case of vanishing cosmological constant~\cite{Perez:2021abf}, where it was shown that, from the point of
view of the asymptotic symmetries, this theory has some unusual properties.
In particular, the asymptotic symmetry algebra does not contain a
generator associated with time translations at the boundary, i.e.,
there is no a notion of energy in this theory\footnote{When Regge-Teitelboim~\cite{Teitelboim:1972vw} and
 ``restricted Henneaux-Troessaert parity conditions'' \cite{Henneaux:2018cst} are used, the asymptotic symmetry
  algebra of the electric theory with $\Lambda=0$ does not contain energy and boost generators~\cite{Perez:2021abf}.
   The same conclusion is obtained when ``extended Henneaux-Troessaert parity conditions''~\cite{Henneaux:2018hdj,Henneaux:2019yax} are
    considered~\cite{Fuentealba:2022gdx}.}.
This effect was also observed in ref.~\cite{Hansen:2021fxi} using
covariant methods. 

When a negative cosmological constant is present, the situation is
even more dramatic. As it was pointed out in~\cite{Perez:2021abf}, the background
configuration~\eqref{eq:background} is no longer a solution of the
electric theory, and as a consequence it does not seems to be possible
to construct a consistent set of asymptotic conditions with non-trivial
charges. The incompatibility of the electric contraction with a negative
cosmological constant was also discussed in~\cite{Hansen:2021fxi}
using scaling arguments.

Here, it is shown that some of these problems can be circumvented
if asymptotic conditions are constructed for the theory that is obtained
from the electric contraction of \emph{Euclidean} General Relativity
with $\Lambda<0$. 

\subsection{Electric theory obtained from a Carrollian contraction of General
Relativity with Lorentzian signature\label{subsec:Electric-theory-obtained}}

The action principle of the theory obtained from an electric contraction
of (Lorentzian) General Relativity with a non-vanishing cosmological
constant is given by
\begin{equation}
I_{\text{Lor}}^{E}=\int dtd^{3}x\left(\pi^{ij}\dot{g}_{ij}-N\mathcal{H}_{\text{Lor}}^{E}-N^{i}\mathcal{H}_{i}^{E}\right)\,,\label{eq:actionelectric}
\end{equation}
where
\begin{align}
\mathcal{H}_{\text{Lor}}^{E} & =\frac{1}{\sqrt{g}}\left(\pi^{ij}\pi_{ij}-\frac{1}{2}\pi^{2}\right)+2\sqrt{g}\Lambda\,,\qquad\qquad\mathcal{H}_{i}^{E}=-2\pi_{i\mid j}^{\;j}\,.\label{eq:constraints_electric}
\end{align}
The Hamiltonian constraint $\mathcal{H}_{\text{Lor}}^{E}$ does not
dependent on the derivatives of the spatial metric. Therefore, as
it was shown in ref.~\cite{Teitelboim:1978wv}, it obeys the following
Poisson bracket
\[
\left\{ \mathcal{H}_{\text{Lor}}^{E}\left(x\right),\mathcal{H}_{\text{Lor}}^{E}\left(x'\right)\right\} =0\,.
\]
This abelian subalgebra reflects the Carrollian structure of the theory~\cite{Henneaux:2021yzg}. The Poisson brackets involving $\mathcal{H}_{i}^{E}$
are completely determined by the invariance of the theory under spatial
reparametrizations, and coincide with those in eqs.~\eqref{eq:HpHi}
and~\eqref{eq:HiHj}.

\subsubsection{Problem with the ground state and the lack of a consistent set of
asymptotic conditions}

From the form of the Hamiltonian constraint $\mathcal{H}_{\text{Lor}}^{E}$
in eq.~\eqref{eq:constraints_electric}, it is clear that the solution
obtained from a naive Carrollian limit of AdS spacetime given in eq.~\eqref{eq:background} is not a solution of this theory. The Hamiltonian
constraint does not admit solutions with vanishing momenta, which
is a direct consequence of the presence of the cosmological constant.
Indeed, in the case when $\Lambda=0$ studied in~\cite{Perez:2021abf},
the direct Carrollian contraction of Minkowski spacetime is a solution
of both Carrollian gravitational theories, electric and magnetic. One could expect
that the background that is used to construct asymptotic symmetries 
possesses some isometries, for example invariance under rotations (that are inherited 
to the asymptotic symmetry algebra). However, as it is shown in
Appendix~\ref{sec:Appendix_A}, there are no spherically symmetric solutions of this theory where the spatial metric
possesses an Euclidean signature, which is one of the main assumptions in the Hamiltonian formulation described here.

The absence of a sensible background configuration has relevant consequences in the possible construction
of a consistent set of asymptotic conditions (usually they are defined
as deviations from the corresponding background). Indeed, if one attempts
to construct asymptotic conditions with finite non-trivial charges
and such that the leading term of the angular components of $g_{ij}$
goes like the metric of a 2-sphere of radius $r$, i.e., $g_{AB}=r^{2}\gamma_{AB}+\dots$,
the Hamiltonian constraint, together with the preservation of the
fall-off, eliminate all the possible terms appearing in the charges,
or fix them as complex numbers.
 This incompatibility with a negative
$\Lambda$ was also observed in ref.~\cite{Hansen:2021fxi} where
the electric Carroll theory was considered as the leading order in
a Carrollian expansion of General Relativity. 

In the next section, it is shown that some of these problems can be
solved if one considers the gravitational Carrollian theory obtained
from the electrical contraction of General Relativity with Euclidean
signature.

\subsection{Electric theory obtained from a Carrollian contraction of General
Relativity with Euclidean signature\label{subsec:Electric-theory-Euclidean}}

The action principle of the electric Carrollian contraction of Euclidean
General Relativity is given by
\begin{equation}
I_{\text{Euc}}^{E}=\int dtd^{3}x\left(\pi^{ij}\dot{g}_{ij}-N\mathcal{H}_{\text{Euc}}^{E}-N^{i}\mathcal{H}_{i}^{E}\right)\,,\label{eq:actionelectric-1}
\end{equation}
where
\begin{align}
\mathcal{H}_{\text{Euc}}^{E} & =-\frac{1}{\sqrt{g}}\left(\pi^{ij}\pi_{ij}-\frac{1}{2}\pi^{2}\right)+2\sqrt{g}\Lambda\,,\qquad\qquad\mathcal{H}_{i}^{E}=-2\pi_{i\mid j}^{\;j}\,.\label{eq:constraints_electric-1}
\end{align}
In this case, there is a sign difference in the first term of the
Hamiltonian constraint, as compared with the one derived from the
Lorentzian theory in eq.~\eqref{eq:constraints_electric}. This is
due the fact that in the Euclidean continuation in General Relativity
one must Wick rotate the canonical momenta $\pi^{ij}\rightarrow-i\pi^{ij}$.
The algebra of the constraints is insensitive to this change of sign,
and consequently it closes according to the algebra in eqs.~\eqref{eq:HpHp}-\eqref{eq:HiHj}.

The transformation laws of the fields are given by
\begin{equation}
\delta g_{ij}=-\frac{2\xi}{\sqrt{g}}\left(\pi_{ij}-\frac{1}{2}g_{ij}\pi\right)+\xi_{i\mid j}+\xi_{j\mid i}\,,\label{eq:transfEl}
\end{equation}

\begin{eqnarray}
\delta\pi^{ij} & = & -\frac{\xi}{2\sqrt{g}}g^{ij}\left(\pi^{kl}\pi_{kl}-\frac{1}{2}\pi^{2}\right)+\frac{2\xi}{\sqrt{g}}\left(\pi_{\;l}^{i}\pi^{jl}-\frac{1}{2}\pi^{ij}\pi\right)-\xi\Lambda\sqrt{g}g^{ij}\nonumber \\
 &  & +\left(\xi^{k}\pi^{ij}\right)_{\mid k}-\xi_{\mid k}^{i}\pi^{kj}-\xi_{\mid k}^{j}\pi^{ki}\,.\label{eq:TransfEl2}
\end{eqnarray}
The variation of the charge then becomes

\begin{equation}
\delta Q_{E}=\oint d^{2}s_{l}\left[2\xi_{k}\delta\pi^{kl}+\left(2\xi^{k}\pi^{jl}-\xi^{l}\pi^{jk}\right)\delta g_{jk}\right]\,.\label{eq:deltaQ-1}
\end{equation}
Note that only the boundary term associated with the momentum constraint,
$\mathcal{H}_{i}^{E}$, contributes to it. 

\subsubsection{Alternative ground state}

For the electric theory obtained from a Carrollian contraction of
Euclidean General Relativity with $\Lambda<0$, it is possible to
find an alternative ground state configuration with a spatial metric
of constant curvature that solves the constraints~\eqref{eq:constraints_electric-1}.
The canonical variables of this solution are given by
\begin{equation}
\bar{g}_{ij}dx^{i}dx^{j}=\frac{dr^{2}}{\left(\frac{r^{2}}{l^{2}}+1\right)}+r^{2}\gamma_{AB}dx^{A}dx^{B}\,,\qquad\bar{\pi}^{ij}=\frac{2}{l}\sqrt{\bar{g}}\bar{g}^{ij}\,,\label{eq:ground-elec}
\end{equation}
with the following lapse and shift
\[
\bar{N}=\sqrt{\frac{r^{2}}{l^{2}}+1}\,,\qquad\bar{N}^{r}=-\frac{r}{l}\sqrt{\frac{r^{2}}{l^{2}}+1}\,.
\]
In contrast with the ground state of the magnetic theory in eq.~\eqref{eq:background},
this configuration has non-trivial momenta and a non-vanishing shift.

The Carrollian Killing vectors can be obtained from the transformations
laws of the canonical variables in eqs.~\eqref{eq:transfEl} and~\eqref{eq:TransfEl2},
requiring that $\delta\bar{g}_{ij}=\delta\bar{\pi}^{ij}=0$. Thus,
they are restricted to obey the following equations

\begin{align}
\xi_{i\mid j}+\xi_{j\mid i}-\frac{2}{3}g_{ij}\xi_{\,\mid k}^{k} & =0\,,\label{eq:CKE}
\end{align}
\begin{equation}
\xi=-\frac{l}{3}\xi_{\,\mid k}^{k}\,.\label{eq:xiCKE}
\end{equation}
Note that eq.~\eqref{eq:CKE} is the three-dimensional conformal Killing
equation defined on a space with Euclidean signature. Because the
composition law~\eqref{eq:comp_Carroll} of $\xi^{i}$ coincides with
the Lie bracket, the algebra of isometries span the conformal group
of the Euclidean three-dimensional space $SO\left(1,4\right)$. 
 As it is shown in the next
section, this property also extends to the asymptotic symmetries that
are defined in terms of deviations with respect to this background.

\subsubsection{Asymptotic conditions}

A consistent set of asymptotic conditions can be constructed by considering
deviations with respect to the background configuration~\eqref{eq:ground-elec}.
The proposed fall-off of the fields is given by
\begin{align}
g_{rr} & =\frac{l^{2}}{r^{2}}-\frac{l^{4}}{r^{4}}+\frac{f_{rr}}{r^{5}}+\mathcal{O}\left(r^{-6}\right)\,,\label{eq:grrelec}\\
g_{rA} & =\frac{f_{rA}}{r^{4}}+\mathcal{O}\left(r^{-5}\right)\,,\\
g_{AB} & =r^{2}\gamma_{AB}+\frac{f_{AB}}{r}+\mathcal{O}\left(r^{-2}\right)\,,
\end{align}
\begin{align}
\pi^{rr} & =\frac{2\sqrt{\gamma}}{l^{2}}r^{3}+\sqrt{\gamma}r+\frac{\sqrt{\gamma}}{l^{2}}\left(\tilde{f}+\frac{1}{l^{2}}f_{rr}\right)+\frac{p^{rr}}{r}+\mathcal{O}\left(r^{-2}\right)\,,\\
\pi^{rA} & =\frac{p^{rA}}{r^{2}}+\mathcal{O}\left(r^{-3}\right)\,,\\
\pi^{AB} & =\frac{2\sqrt{\gamma}\gamma^{AB}}{r}-\frac{\sqrt{\gamma}\gamma^{AB}l^{2}}{r^{3}}-\frac{\sqrt{\gamma}\tilde{f}^{AB}}{2r^{4}}+\frac{p^{AB}}{r^{5}}+\mathcal{O}\left(r^{-6}\right)\,.\label{eq:piABelec}
\end{align}
The asymptotic behavior of the canonical variables is preserved by
the following parameters
\begin{align}
\xi & =r\,T+\frac{l^{2}}{4}\left(\Delta+2\right)T\frac{1}{r}+\dots\,,\\
\xi^{r} & =-\frac{r^{2}T}{l}-\frac{rD_{A}Y^{A}}{2}+\frac{l}{4}\left(\Delta-2\right)T+\dots\,,\label{eq:xirelec}\\
\xi^{A} & =Y^{A}-\left(D^{A}T\right)\frac{l}{r}+\dots\,.\label{eq:xiAelec}
\end{align}
Here, the parameters $T$ and $Y^{A}$ must to obey the following
conditions coming from the preservation of~\eqref{eq:grrelec}-\eqref{eq:piABelec}
\begin{equation}
D_{A}D_{B}T-\frac{1}{2}\gamma_{AB}\Delta T=0\,,\label{eq:Param1}
\end{equation}
\begin{equation}
D_{A}Y_{B}+D_{B}Y_{A}-\gamma_{AB}D_{C}Y^{C}=0\,.\label{eq:ParamYElec}
\end{equation}

The conserved charges can be obtained using the asymptotic conditions~\eqref{eq:grrelec}-\eqref{eq:xiAelec} in eq.~\eqref{eq:deltaQ-1}
\begin{equation}
Q=\oint d^{2}x\left(-\frac{4\sqrt{\gamma}}{l^{3}}Tf_{rr}+2Y^{A}\gamma_{AB}p^{rB}\right)\,,\label{eq:QELEct1}
\end{equation}
where it was assumed that the parameters $T$ and $Y^{A}$ have a
vanishing functional variation. Note that, in spite of the fact that
according to eq.~\eqref{eq:deltaQ-1} the parameter $\xi$ describing
normal deformations to the $t=const.$ hypersurface does not appear
in $\delta Q$, the parameter $T$ in the leading order of $\xi$
enters in the expression for the charge through $\xi^{r}$. There
is no analogue of this situation when the cosmological constant vanishes
\cite{Perez:2021abf}.

The explicit solution of eqs.~\eqref{eq:Param1} and~\eqref{eq:ParamYElec}
is

\begin{align*}
T & =\alpha^{40}+\alpha^{4I}\hat{r}_{I}\,,\\
Y^{A} & =\partial^{A}\left(\alpha^{0I}\hat{r}_{I}\right)+\frac{\epsilon^{AB}}{\sqrt{\gamma}}\partial_{B}\left(\frac{1}{2}\epsilon_{IJK}\alpha^{JK}\hat{r}_{I}\right)\,,
\end{align*}
where $\alpha^{40}$, $\alpha^{4I}$, $\alpha^{0I}$ and $\alpha^{JK}$
are constants in the angles. Therefore, the charge $Q$ in~\eqref{eq:QELEct1}
can be written as
\[
Q=\frac{1}{2}\alpha^{mn}J_{mn}\,,
\]
where $m,n=\{0,I,4\}$, with $I=1,2,3$ and $\alpha_{mn}=-\alpha_{nm}$,
$J_{mn}=-J_{nm}$.

The explicit form of the generators expressed in terms of surface
integrals is 
\[
J_{40}=-\oint d^{2}x\left(\frac{4\sqrt{\gamma}}{l^{3}}f_{rr}\right)\,,\qquad J_{4I}=-\oint d^{2}x\left(\frac{4\sqrt{\gamma}}{l^{3}}\hat{r}_{I}f_{rr}\right)\,,
\]
\[
J_{0I}=-\oint d^{2}x\left(2\hat{r}_{I}D_{A}p^{rA}\right)\,,\qquad J_{IJ}=\oint d^{2}x\left(2\sqrt{\gamma}\epsilon_{IJK}\epsilon_{AB}\hat{r}_{K}D^{A}p^{rB}\right)\,.
\]
The asymptotic symmetry algebra can then be obtained using the identity
$\delta_{2}Q\left[\lambda_{1}\right]=\left\{ Q\left[\lambda_{1}\right],Q\left[\lambda_{2}\right]\right\} $,
where the transformation laws of the fields appearing in the charges,
$f_{rr}$ and $p^{rA}$, are given by 

\[
\delta f_{rr}=\frac{3}{2}f_{rr}D_{A}Y^{A}+Y^{A}D_{A}f_{rr}+\frac{l^{3}}{2\sqrt{\gamma}}TD_{A}p^{rA}+l^{3}\left(D_{A}T\right)\frac{p^{rA}}{\sqrt{\gamma}}\,,
\]

\begin{align*}
\delta p^{rA} & =-\frac{3\sqrt{\gamma}}{l^{3}}f_{rr}D^{A}T-\frac{\sqrt{\gamma}}{l^{3}}TD^{A}f_{rr}+\frac{3\sqrt{\gamma}}{2l}TD_{C}\tilde{f}^{AC}+\frac{3\sqrt{\gamma}}{2l}\left(D_{B}T\right)\tilde{f}^{AB}\\
 & \;\;\;+\left(D_{B}Y^{B}\right)p^{rA}+\mathcal{L}_{Y}p^{rA}\,.
\end{align*}
The generators $J_{mn}$ then obey the following Poisson bracket algebra
\[
\left\{ J_{mn},J_{rs}\right\} =\eta_{ms}J_{nr}+\eta_{nr}J_{ms}-\eta_{mr}J_{ns}-\eta_{ns}J_{mr}\,,
\]
where $\eta_{mn}=\text{diag}\left(-,+,+,+,+\right)$. Thus, the asymptotic
symmetry algebra is given the $so\left(1,4\right)$ algebra. 

Although this symmetry is not the Carroll algebra (or a cosmological extension of it), it could be interpreted as the Carrollian symmetry 
that acts on the homogeneous space given by the  ``Carrollian light-cone''~\cite{Figueroa-OFarrill:2019sex}. 

It is worth noting that the asymptotic conditions~\eqref{eq:grrelec}-\eqref{eq:piABelec}
are not inherited from General Relativity, and therefore are specific
of this particular electric Carrollian theory.

\subsubsection{Spherically symmetric solutions}

As it was shown in the previous sections, if one considers the Carrollian
theory obtained from an electric contraction of Euclidean General
Relativity with $\Lambda<0$, a consistent set of asymptotic conditions
can be constructed using the ground state~\eqref{eq:ground-elec}
as a starting point. This does not seem to be possible for the Carrollian
theory obtained from Einstein gravity with a Lorentzian signature.

In spite of this success, the electric Carrollian theory coming from
Euclidean General Relativity still possesses some unusual properties
that become explicit when spherically symmetric solutions are considered.
For example, the spherically symmetric ansatz of the form
\begin{equation}
g_{ij}dx^{i}dx^{j}=\frac{dr^{2}}{f^{2}\left(r\right)}+r^{2}\left(d\theta^{2}+\sin^{2}\theta d\phi^{2}\right)\,,\qquad\pi^{ij}=\frac{2}{l}\sqrt{g}g^{ij}\,,\label{eq:spherically symmans}
\end{equation}
solves the equation of motion for any arbitrary value of $f\left(r\right)$,
with the following lapse and shift 
\[
N=f\left(r\right)\qquad,\qquad N^{r}=-\frac{r}{l}f\left(r\right)\,.
\]
Consequently, there is a huge degeneracy in the space of solutions
of this theory. Similar problems were encountered in ref.~\cite{Perez:2021abf}
when $\Lambda=0$. 

If we choose $f\left(r\right)$ such that the spatial metric in~\eqref{eq:spherically symmans}
coincides with the one of the (Euclidean) Schwarzschild-AdS metric
(that contains the ground state configuration~\eqref{eq:ground-elec}
as a particular case), i.e.,
\[
f\left(r\right)=\sqrt{\frac{r^{2}}{l^{2}}+1-\frac{M}{8\pi r}}\,,
\]
the only non-trivial charge is $J_{40}=-2lM$, which is proportional
to the parameter $M$ that in General Relativity represents the ``mass''
of the solution. 

\section{Discussion\label{sec:5 Final-remarks}}

The asymptotic structure of Carrollian gravitational theories, obtained
from electric and magnetic contractions of General Relativity with
a negative cosmological constant, was analyzed. The electric theory,
as in the case with $\Lambda=0$, has some unusual features from
the point of view of their asymptotic symmetries. In particular, for
the electric theory obtained from a contraction of (Lorentzian) General
Relativity, it does not seems to be possible to construct a consistent
set of asymptotic conditions. This can be traced back to the absence
of a sensible ground state configuration in this theory. The situation
can be improved if the Carrollian theory obtained from an electric
contraction of Euclidean General Relativity is considered. In this
case, asymptotic conditions are constructed with an asymptotic symmetry
algebra given by the ``Euclidean AdS$_{4}$'' algebra $so\left(1,4\right)$. This feature has
its origin in the fact that the fall-off of the fields cannot be seen
as coming from Einstein gravity, it is intrinsic to the ``Euclidean''
electric Carrollian theory. Additionally, it was shown that the space
of a certain class of spherically symmetric solutions is degenerate, where some of the arbitrary 
functions in the ansatz are not determined by the equations of motion.

On the contrary, the magnetic Carrollian gravitational theory with
$\Lambda<0$ has very interesting properties and a very rich structure
from the perspective of their asymptotic symmetries. The asymptotic
symmetry algebra is infinite-dimensional
and corresponds to the three-dimensional Carrollian conformal algebra,
that according to~\cite{Duval:2014uva} is isomorphic to the BMS$_{4}$
algebra. These results are in full agreement with the expectations
coming from holography. In this sense, this theory could provide a
new concrete framework for the study of Carrollian holography. This
is particularly interesting because the structure of the constraints
in the Carrollian magnetic theory is simpler than the one of General
Relativity. Indeed, the algebra~\eqref{eq:HpHp}-\eqref{eq:HiHj}
is field independent. Furthermore, as it is shown in ref.~\cite{Perez-Troncoso-2022},
this theory also admits a solution (see eqs.~\eqref{eq:mag_Schw_AdS}
and~\eqref{eq:lsmag_Schw_AdS}) that can be interpreted as a ``Carrollian
black hole.'' It is a thermal regular configuration in the context
of Carrollian geometry, that possesses a non-vanishing entropy. This
could provide a fertile ground to explore the possible nature of their
microstates, as well as different aspects related with quantum gravity
and holography in the Carrollian limit.

\acknowledgments{I wish to thank Oscar Fuentealba, Marc Henneaux, Stefan Prohazka, Jakob Salzer and Ricardo Troncoso for useful discussions and comments. I also thank the organizers of the ``Carroll Workshop'' in Vienna, where this work was completed. This research has been partially supported by Fondecyt grants N$\textsuperscript{\underline{o}}$ 1211226, 1220910.}

\appendix
\section{Spherically symmetric configurations in ``Lorentzian'' electric Carroll gravity with a negative cosmological constant }
\label{sec:Appendix_A}
In this appendix we examine spherically symmetric configuration in
the electric Carrollian contraction of Lorentzian Einstein gravity
with a negative cosmological constant. 

Let us consider a spherically symmetric ansatz for the canonical variables
of the following form
\[
g_{rr}=f\left(r\right)\,,\qquad g_{rA}=0\,,\qquad g_{AB}=r^{2}\gamma_{AB}\,,
\]
\[
\pi^{rr}=\sqrt{g}p\left(r\right)\,,\qquad\pi^{rA}=0\,,\qquad\pi^{AB}=\sqrt{\gamma}\gamma^{AB}h\left(r\right)\,,
\]
From the constraints $\mathcal{H}_{\text{Lor}}^{E}\approx0$ and $\mathcal{H}_{i}^{E}\approx0$
in eq.~\eqref{eq:constraints_electric} one finds
\[
f\left(r\right)=\frac{r}{p\left(r\right)^{2}}\left(C-\frac{4r^{3}}{\ell^{2}}\right)\,,\qquad h\left(r\right)=\frac{1}{p\left(r\right)}\left(\frac{C}{4r}-\frac{4r^{2}}{\ell^{2}}\right)\,,
\]
where $C$ is an integration constant. Note that for those values
of $r$ such that $r>\left(\frac{\ell^{2}C}{4}\right)^{\frac{1}{3}}$,
the function $f\left(r\right)$ becomes negative and therefore violates
the assumption that the spatial metric has Euclidean signature. In
particular, this is always true in the asymptotic region where $r\rightarrow\infty$. 

As a consequence, spherically symmetric configurations with an Euclidean spatial metric are not
solutions of the electric Carrollian gravity with negative cosmological constant
that is obtained from the ultra-relativistic limit of Lorentzian General
Relativity.

\bibliographystyle{JHEP}
\bibliography{review}

\end{document}